\documentstyle[epsf,epsfig,here,12pt]{article}  
\begin{document}

\newcommand{\rhobar}{\overline{\rho}}
\newcommand{\etabar}{\overline{\eta}}
\newcommand{\epsilonk}{\left | \epsilon_K \right |}
\newcommand{\vubovcb}{\left | \frac{V_{ub}}{V_{cb}} \right |}
\newcommand{\vtdovts}{\left | \frac{V_{td}}{V_{ts}} \right |}
\newcommand{\dmd}{\Delta m_d}
\newcommand{\dms}{\Delta m_s}
\newcommand{\pr}{{\rm P.R.}}
\newcommand{\Ds}{{\rm D}_s^+}
\newcommand{\Dp}{{\rm D}^+}
\newcommand{\Do}{{\rm D}^0}
\newcommand{\piss}{\pi^{\ast \ast}}
\newcommand{\pis}{\pi^{\ast}}
\newcommand{\bbar}{\overline{b}}
\newcommand{\cbar}{\overline{c}}
\newcommand{\Dstar}{{\rm D}^{\ast}}
\newcommand{\Dstars}{{\rm D}^{\ast +}_s}
\newcommand{\Dstaro}{{\rm D}^{\ast 0}}
\newcommand{\Dstarstar}{{\rm D}^{\ast \ast}}
\newcommand{\Dbar}{\overline{{\rm D}}}
\newcommand{\Bbar}{\overline{{\rm B}}}
\newcommand{\Bsbar}{\overline{{\rm B}^0_s}}
\newcommand{\Lcbar}{\overline{\Lambda^+_c}}
\newcommand{\nubar}{\overline{\nu_{\ell}}}
\newcommand{\tautaubar}{\tau \overline{\tau}}
\newcommand{\Vcb}{\left | {\rm V}_{cb} \right |}
\newcommand{\Vub}{\left | {\rm V}_{ub} \right |}
\newcommand{\Vtd}{\left | {\rm V}_{td} \right |}
\newcommand{\Vts}{\left | {\rm V}_{ts} \right |}
\newcommand{\fleisher}{\frac{BR({\rm B}^0~(\overline{{\rm B}^0}) \rightarrow \pi^{\pm} {\rm K}^{\mp})}
{BR({\rm B}^{\pm} \rightarrow \pi^{\pm} {\rm K}^0)}}
\newcommand{\bptre}{\rm b^{+}_{3}}
\newcommand{\bp}{\rm b^{+}_{1}}
\newcommand{\bo}{\rm b^0}
\newcommand{\bos}{\rm b^0_s}
\newcommand{\bss}{\rm b^s_s}
\newcommand{\qq}{\rm q \overline{q}}
\newcommand{\cc}{\rm c \overline{c}}
\newcommand{\BsDmX}{{B_{s}^{0}} \rightarrow D \mu X}
\newcommand{\BsDsm}{{B_{s}^{0}} \rightarrow D_{s} \mu X}
\newcommand{\BsDsX}{{B_{s}^{0}} \rightarrow D_{s} X}
\newcommand{\BDsX}{B \rightarrow D_{s} X}
\newcommand{\BDomX}{B \rightarrow D^{0} \mu X}
\newcommand{\BDpmX}{B \rightarrow D^{+} \mu X}
\newcommand{\Dsfmn}{D_{s} \rightarrow \phi \mu \nu}
\newcommand{\Dsfipi}{D_{s} \rightarrow \phi \pi}
\newcommand{\DsfX}{D_{s} \rightarrow \phi X}
\newcommand{\DpfX}{D^{+} \rightarrow \phi X}
\newcommand{\DofX}{D^{0} \rightarrow \phi X}
\newcommand{\DfX}{D \rightarrow \phi X}
\newcommand{\DsD}{B \rightarrow D_{s} D}
\newcommand{\DsmX}{D_{s} \rightarrow \mu X}
\newcommand{\DmX}{D \rightarrow \mu X}
\newcommand{\Zbb}{Z^{0} \rightarrow \rm b \overline{b}}
\newcommand{\Zcc}{Z^{0} \rightarrow \rm c \overline{c}}
\newcommand{\Rbb}{\frac{\Gamma_{Z^0 \rightarrow \rm b \overline{b}}}
{\Gamma_{Z^0 \rightarrow Hadrons}}}
\newcommand{\Rcc}{\frac{\Gamma_{Z^0 \rightarrow \rm c \overline{c}}}
{\Gamma_{Z^0 \rightarrow Hadrons}}}
\newcommand{\bb}{\rm b \overline{b}}
\newcommand{\str}{\rm s \overline{s}}
\newcommand{\Bs}{\rm{B^0_s}}
\newcommand{\Bsb}{\overline{\rm{B^0_s}}}
\newcommand{\Bp}{\rm{B^{+}}}
\newcommand{\Bm}{\rm{B^{-}}}
\newcommand{\Bo}{\rm{B^{0}}}
\newcommand{\Bd}{\rm{B^{0}_{d}}}
\newcommand{\Bdb}{\overline{\rm{B^{0}_{d}}}}
\newcommand{\Lb}{\Lambda^0_b}
\newcommand{\Lbb}{\overline{\Lambda^0_b}}
\newcommand{\Kstar}{\rm{K^{\star 0}}}
\newcommand{\phim}{\rm{\phi}}
\newcommand{\Dsp}{\mbox{D}_s^+}
\newcommand{\Dn}{\mbox{D}^0}
\newcommand{\Dsb}{\overline{\mbox{D}_s}}
\newcommand{\Dm}{\mbox{D}^-}
\newcommand{\Dnb}{\overline{\mbox{D}^0}}
\newcommand{\Lc}{\Lambda_c}
\newcommand{\Lcb}{\overline{\Lambda_c}}
\newcommand{\Dstarp}{\mbox{D}^{\ast +}}
\newcommand{\Dstarm}{\mbox{D}^{\ast -}}
\newcommand{\Dsstarp}{\mbox{D}_s^{\ast +}}
\newcommand{\Km}{\mbox{K}^-}
\newcommand{\Pb}{P_{b-baryon}}
\newcommand{\KKpi}{\rm{ K K \pi }}
\newcommand{\GeV}{\rm{GeV}}
\newcommand{\MeV}{\rm{MeV}}
\newcommand{\nb}{\rm{nb}}
\newcommand{\Zzero}{{\rm Z}^0}
\newcommand{\MZ}{\rm{M_Z}}
\newcommand{\MW}{\rm{M_W}}
\newcommand{\GF}{\rm{G_F}}
\newcommand{\Gm}{\rm{G_{\mu}}}
\newcommand{\MH}{\rm{M_H}}
\newcommand{\MT}{\rm{m_{top}}}
\newcommand{\GZ}{\Gamma_{\rm Z}}
\newcommand{\Afb}{\rm{A_{FB}}}
\newcommand{\Afbs}{\rm{A_{FB}^{s}}}
\newcommand{\sigmaf}{\sigma_{\rm{F}}}
\newcommand{\sigmab}{\sigma_{\rm{B}}}
\newcommand{\NF}{\rm{N_{F}}}
\newcommand{\NB}{\rm{N_{B}}}
\newcommand{\Nnu}{\rm{N_{\nu}}}
\newcommand{\RZ}{\rm{R_Z}}
\newcommand{\rhob}{\rho_{eff}}
\newcommand{\Gammanz}{\rm{\Gamma_{Z}^{new}}}
\newcommand{\Gammani}{\rm{\Gamma_{inv}^{new}}}
\newcommand{\Gammasz}{\rm{\Gamma_{Z}^{SM}}}
\newcommand{\Gammasi}{\rm{\Gamma_{inv}^{SM}}}
\newcommand{\Gammaxz}{\rm{\Gamma_{Z}^{exp}}}
\newcommand{\Gammaxi}{\rm{\Gamma_{inv}^{exp}}}
\newcommand{\rhoZ}{\rho_{\rm Z}}
\newcommand{\thw}{\theta_{\rm W}}
\newcommand{\swsq}{\sin^2\!\thw}
\newcommand{\swsqmsb}{\sin^2\!\theta_{\rm W}^{\overline{\rm MS}}}
\newcommand{\swsqbar}{\sin^2\!\overline{\theta}_{\rm W}}
\newcommand{\cwsqbar}{\cos^2\!\overline{\theta}_{\rm W}}
\newcommand{\swsqb}{\sin^2\!\theta^{eff}_{\rm W}}
\newcommand{\ee}{{e^+e^-}}
\newcommand{\eeX}{{e^+e^-X}}
\newcommand{\gaga}{{\gamma\gamma}}
\newcommand{\mumu}{\ifmmode {\mu^+\mu^-} \else ${\mu^+\mu^-} $ \fi}
\newcommand{\eeg}{{e^+e^-\gamma}}
\newcommand{\mumug}{{\mu^+\mu^-\gamma}}
\newcommand{\tautau}{{\tau^+\tau^-}}
\newcommand{\qqb}{{q\bar{q}}}
\newcommand{\eegg}{e^+e^-\rightarrow \gamma\gamma}
\newcommand{\eeggg}{e^+e^-\rightarrow \gamma\gamma\gamma}
\newcommand{\eeee}{e^+e^-\rightarrow e^+e^-}
\newcommand{\eeeeee}{e^+e^-\rightarrow e^+e^-e^+e^-}
\newcommand{\eeeeg}{e^+e^-\rightarrow e^+e^-(\gamma)}
\newcommand{\eeeegg}{e^+e^-\rightarrow e^+e^-\gamma\gamma}
\newcommand{\eeeg}{e^+e^-\rightarrow (e^+)e^-\gamma}
\newcommand{\eemumu}{e^+e^-\rightarrow \mu^+\mu^-}
\newcommand{\eetautau}{e^+e^-\rightarrow \tau^+\tau^-}
\newcommand{\eehad}{e^+e^-\rightarrow {\rm hadrons}}
\newcommand{\eettg}{e^+e^-\rightarrow \tau^+\tau^-\gamma}
\newcommand{\eell}{e^+e^-\rightarrow l^+l^-}
\newcommand{\Ztopig}{{\rm Z}^0\rightarrow \pi^0\gamma}
\newcommand{\Ztogg}{{\rm Z}^0\rightarrow \gamma\gamma}
\newcommand{\Ztoee}{{\rm Z}^0\rightarrow e^+e^-}
\newcommand{\Ztoggg}{{\rm Z}^0\rightarrow \gamma\gamma\gamma}
\newcommand{\Ztomumu}{{\rm Z}^0\rightarrow \mu^+\mu^-}
\newcommand{\Ztotautau}{{\rm Z}^0\rightarrow \tau^+\tau^-}
\newcommand{\Ztoll}{{\rm Z}^0\rightarrow l^+l^-}
\newcommand{\Ztocc}{{\rm Z^0\rightarrow c \bar c}}
\newcommand{\Lamp}{\Lambda_{+}}
\newcommand{\Lamm}{\Lambda_{-}}
\newcommand{\Pt}{\rm P_{t}}
\newcommand{\Gee}{\Gamma_{ee}}
\newcommand{\Gpig}{\Gamma_{\pi^0\gamma}}
\newcommand{\Ggg}{\Gamma_{\gamma\gamma}}
\newcommand{\Gggg}{\Gamma_{\gamma\gamma\gamma}}
\newcommand{\Gmumu}{\Gamma_{\mu\mu}}
\newcommand{\Gtautau}{\Gamma_{\tau\tau}}
\newcommand{\Ginv}{\Gamma_{\rm inv}}
\newcommand{\Ghad}{\Gamma_{\rm had}}
\newcommand{\Gnu}{\Gamma_{\nu}}
\newcommand{\GnuSM}{\Gamma_{\nu}^{\rm SM}}
\newcommand{\Gll}{\Gamma_{l^+l^-}}
\newcommand{\Gff}{\Gamma_{f\overline{f}}}
\newcommand{\Gtot}{\Gamma_{\rm tot}}
\newcommand{\Rb}{\mbox{R}_b}
\newcommand{\Rc}{\mbox{R}_c}
\newcommand{\al}{a_l}
\newcommand{\vl}{v_l}
\newcommand{\af}{a_f}
\newcommand{\vf}{v_f}
\newcommand{\ael}{a_e}
\newcommand{\ve}{v_e}
\newcommand{\amu}{a_\mu}
\newcommand{\vmu}{v_\mu}
\newcommand{\atau}{a_\tau}
\newcommand{\vtau}{v_\tau}
\newcommand{\ahatl}{\hat{a}_l}
\newcommand{\vhatl}{\hat{v}_l}
\newcommand{\ahate}{\hat{a}_e}
\newcommand{\vhate}{\hat{v}_e}
\newcommand{\ahatmu}{\hat{a}_\mu}
\newcommand{\vhatmu}{\hat{v}_\mu}
\newcommand{\ahattau}{\hat{a}_\tau}
\newcommand{\vhattau}{\hat{v}_\tau}
\newcommand{\vtildel}{\tilde{\rm v}_l}
\newcommand{\avsq}{\ahatl^2\vhatl^2}
\newcommand{\Ahatl}{\hat{A}_l}
\newcommand{\Vhatl}{\hat{V}_l}
\newcommand{\Afer}{A_f}
\newcommand{\Ael}{A_e}
\newcommand{\Aferb}{\bar{A_f}}
\newcommand{\Aelb}{\bar{A_e}}
\newcommand{\AVsq}{\Ahatl^2\Vhatl^2}
\newcommand{\Iwk}{I_{3l}}
\newcommand{\Qch}{|Q_{l}|}
\newcommand{\roots}{\sqrt{s}}
\newcommand{\pT}{p_{\rm T}}
\newcommand{\mt}{m_t}
\newcommand{\Rechi}{{\rm Re} \left\{ \chi (s) \right\}}
\newcommand{\up}{^}
\newcommand{\abscosthe}{|cos\theta|}
\newcommand{\dsum}{\Sigma |d_\circ|}
\newcommand{\zsum}{\Sigma z_\circ}
\newcommand{\sint}{\mbox{$\sin\theta$}}
\newcommand{\cost}{\mbox{$\cos\theta$}}
\newcommand{\mcost}{|\cos\theta|}
\newcommand{\epair}{\mbox{$e^{+}e^{-}$}}
\newcommand{\mupair}{\mbox{$\mu^{+}\mu^{-}$}}
\newcommand{\taupair}{\mbox{$\tau^{+}\tau^{-}$}}
\newcommand{\gamgam}{\mbox{$e^{+}e^{-}\rightarrow e^{+}e^{-}\mu^{+}\mu^{-}$}}
\newcommand{\fullskip}{\vskip 16cm}
\newcommand{\halfskip}{\vskip  8cm}
\newcommand{\quarskip}{\vskip  6cm}
\newcommand{\abitskip}{\vskip 0.5cm}
\newcommand{\ba}{\begin{array}}
\newcommand{\ea}{\end{array}}
\newcommand{\bc}{\begin{center}}
\newcommand{\ec}{\end{center}}
\newcommand{\be}{\begin{eqnarray}}
\newcommand{\eeq}{\end{eqnarray}}
\newcommand{\bes}{\begin{eqnarray*}}
\newcommand{\ees}{\end{eqnarray*}}
\newcommand{\Kz}{\ifmmode {\rm K^0_s} \else ${\rm K^0_s} $ \fi}
\newcommand{\Zz}{\ifmmode {\rm Z^0} \else ${\rm Z^0 } $ \fi}
\newcommand{\qqbar}{\ifmmode {\rm q\bar{q}} \else ${\rm q\bar{q}} $ \fi}
\newcommand{\ccbar}{\ifmmode {\rm c\bar{c}} \else ${\rm c\bar{c}} $ \fi}
\newcommand{\bbbar}{\ifmmode {\rm b\bar{b}} \else ${\rm b\bar{b}} $ \fi}
\newcommand{\xxbar}{\ifmmode {\rm x\bar{x}} \else ${\rm x\bar{x}} $ \fi}
\newcommand{\rphi}{\ifmmode {\rm R\phi} \else ${\rm R\phi} $ \fi}
%\renewcommand {\pt}         {\rm p_t}
%========================================================================% 

%################################################## titlepage declaration

\begin{titlepage}

\pagenumbering{arabic}
%\vspace*{-1.5cm}
%\begin{tabular*}{15.cm}{l@{\extracolsep{\fill}}r}
%{  } & 
%%===================> DELPHI note number       =====> To be filled <=====%
%DELPHI 97-XXX PHYS YYY
%%========================================================================%
%\\
%& 
%%===================> DELPHI note date         =====> To be filled <=====%
%17 December, 1997
%%========================================================================%
%\\
%&\\ \hline
%\end{tabular*}
%\vspace*{2.cm}

\begin{center}
\Large 
{\bf Constraints on the parameters of the $V_{CKM}$ matrix
at the end of 1997. } 
\vspace*{1.cm}
\\
\normalsize {    {\bf 
  F. Parodi$^{1,2}$, P. Roudeau$^{1}$ and A. Stocchi$^{1}$ }\\
\vskip 0.5truecm
   {\footnotesize ( (1) Laboratoire de l'Acc\'el\'erateur Lin\'eaire (L.A.L.) - Orsay ) } \\ 
   {\footnotesize ( (2) I.N.F.N. Genova )  } }
\end{center}

\begin{abstract}
\noindent
A review of the current status of the Cabibbo-Kobayashi-Maskawa matrix ($V_{CKM}$) is presented.
This paper contains an update of the 
results published in \cite{ref:bello}. Values of the parameters entering into the constraints, which restrict the range for
$\overline{\rho}$ and $\overline{\eta}$ parameters, include recent measurements given at 
1997 Summer Conferences and progress obtained by lattice QCD collaborations.
Experimental constraints imposed by the measurements of $\epsilonk$, $\vubovcb$, $\dmd$ and
by the limit on $\dms$, are compatible and do not show evidence for New Physics inside
measurements errors. Values for the angles $\alpha$, $\beta$ and $\gamma$ of the C.K.M.
triangle have been also obtained:
$$
\rhobar=0.156 \pm 0.090  ,~\etabar=0.328 \pm 0.054 
$$
$$
\sin 2 \alpha =   -0.10 \pm 0.40    ,~\sin 2 \beta =    0.68 \pm 0.10     ,~\gamma= (64  \pm 12)^{\circ}
$$
Using the analysis of \cite{ref:aligreub}, the ratio of the branching fractions for charged and
neutral B mesons decaying into ${\rm K} \pi$ can be predicted with good precision:
$$
R1=\fleisher= 0.89 \pm 0.08
$$
Angles $ \theta$, $\theta_u$, $\theta_d$ and $\phi$ proposed in the parametrisation \cite{ref:friche}
of the C.K.M. matrix have been also determined:
$$
\theta= (2.30 \pm 0.09 )^{\circ},~\theta_u= (4.87 \pm 0.86)^{\circ}
$$
$$
\theta_d= (11.71 \pm 1.09)^{\circ},~\phi= (91.1 \pm 11.8)^{\circ}
$$
The present value of $\phi$ is compatible with the maximum CP violation scenario advocated in \cite{ref:friche}.

As there are more constraints than the fitted $\rhobar$ and $\etabar$ parameters two studies have been done.

In the first study, several external measurements or theoretical inputs have been removed, in turn, from the 
constraints and their respective values have been fitted simultaneously with $\rhobar$ and $\etabar$.
Central values and uncertainties on these quantities  have been compared with actual measurements
or theoretical evaluations. In this way it is possible to quantify the importance of the different
measurements and the coherence of the Standard Model scenario for CP violation.

In the other study an additional parameter has been introduced to account for New Physics, 
inside a given model and limits on the masses of the new particles appearing in this model
have been updated by reference to the similar analysis done in \cite{ref:bello}.\\

\end{abstract}

%\vskip 2.0truecm
%\vspace{\fill}
%\begin{center}
%%To be submitted to  CERN-OPEN
%\end{center}
%\noindent
%%+~~~~~ on leave of absence of the INFN-Genova
%\vspace{\fill}
\end{titlepage}

\setcounter{page}{1}    

\section {Introduction.}
In a previous publication \cite{ref:bello}, uncertainties on the determination of the C.K.M.
parameters $A$, $\rhobar$ and $\etabar$ corresponding to the Wolfenstein parametrization,
have been reviewed
\footnote{$\rhobar$ and $\etabar$ are related to the original $\rho$ and $\eta$
parameters introduced by L. Wolfenstein: $\rhobar(\etabar)= \rho(\eta)(1-\frac{\lambda^2}{2})$
\cite{ref:burbur}}. 
This study was based on measurements and theoretical estimates available
at the beginning of 1997. Present results are obtained using new measurements and theoretical 
analyses available by summer 1997.

\begin{table}[htb]
\begin{center}
\begin{tabular}{|c|c|}
\hline
 parameters used in \cite{ref:bello} & present analysis \\
\hline
 A  ~=~0.81 $\pm$ 0.04  & $0.823 \pm 0.033$ \\
 $\frac{\left | V_{ub} \right |}{\left | V_{cb} \right |} ~=~0.08 \pm 0.02 $ & unchanged \\
 $\Delta m_d~=~(0.469 \pm 0.019) ~ps^{-1}$  & $~(0.472 \pm 0.018) ~ps^{-1}$  \\
 $\Delta m_s > 8.0 ps^{-1} ~{\rm at}~95 \%~C.L.$ & $>$ 10.2 ps$^{-1} ~{\rm at}~95 \% ~C.L.$ \\
 $ \overline{m_t}(m_t)~=~(168 \pm 6) ~GeV/c^2$  & unchanged \\
 $B_K~=~0.90 \pm 0.09$  & unchanged \\
 $ f_{B_d} \sqrt{B_{B_d}}~=~(200 \pm 50)~MeV$  & $~(220 \pm 40)~MeV$ \\
 $\xi = \frac{ f_{B_s} \sqrt{B_{B_s}}}{ f_{B_d} \sqrt{B_{B_d}}}~=~1.17 \pm 0.13$ & $1.10 \pm 0.07$\\
\hline
\end{tabular}
\caption[]{ \it {Values of the relevant parameters entering into the 
Standard Model expressions for $\Delta m_d$,
$\left | \epsilon_K \right |$, $\frac{\left | V_{ub}\right |}{\left | V_{cb} \right |}$  and $\Delta m_d/\Delta m_s$.
The second column gives the results, updated with respect to the previous publication\cite{ref:bello}, 
which are used in this paper.}}
\label{tab:a}
\end{center}
\end{table}

The central values and uncertainties for the relevant parameters used in this analysis are given in 
Table \ref{tab:a}.

This analysis (see also \cite{ref:bello}) differs from similar studies \cite{ref:concurrence}  because
recent results from experiments and from lattice QCD have been included. 
This is of peculiar importance in the determination
of $A$, $B_K$ and of the ratio $\xi$. The constraint from $\epsilonk$ is then
really effective because it selects a region in $\etabar$ 
of similar size as the one given by the other constraints. This implies that
%, for thefirst time, 
the compatibility between the measurements of the sides of the C.K.M. triangle
(given by the values of $\frac{\left | V_{ub} \right |}{\left | V_{cb} \right |}$, $\dmd$ and 
$\dms$) and a measurement directly related to CP violation (given by the value of $\epsilonk$)
is investigated.

%The central
%values and the uncertainties of certain parameters have chenged with respect to those given in \cite{ref:bello}. The reasons of
%these changes are given in details in the following.
In the following sections, the procedures used to obtain these new values have been explained.
Section \ref{sec:A} details the evaluation of $A$ through the measurement of the $\Vcb$ element
of the C.K.M. matrix. In section \ref{sec:dms} improvements in the determination of other important
parameters in this analysis are considered.
The new limit on $\dms$ obtained by the LEP experiments is recalled. 
%In section \ref{sec:fb}, 
The present determination of $f_B$,
through the measurement of $f_{D_s}$ and the use of lattice QCD is explained \cite{ref:bello}. 
%Finally, in section \ref{sec:fbsofbd}, 
New results from lattice QCD relating the $\Bd$ and $\Bs$ decay 
constants are reported and finally the value used for $B_K$ is commented.

Using the constraints on $\rhobar$ and $\etabar$ provided by the measurements
of $\epsilonk$, $\dmd$, $\dms$ and $\vubovcb$, in the framework of the Standard Model,
the region selected in the $(\rhobar,~\etabar)$ plane and the determination of the 
angles $\alpha$, $\beta$ and $\gamma$ of the unitarity triangle, are analyzed in section
\ref{sec:mesures}.
The parametrization of the C.K.M. matrix proposed in \cite{ref:friche} has been also considered 
and the corresponding parameters have been determined.

As there are more experimental constraints than fitted parameters, the information
coming from one of the external parameters ($\dms$, $m_t$, $A$, $\vubovcb$, $B_K$, and $B_{B_d} \sqrt{f_{B_d}}$)
has been removed, in turn, from the fit and the probability density distribution for this parameter,
determined by the other parameters and constraints has been determined in section \ref{sec:param}.

Another possibility consists in fitting an extra parameter, in addition to $\rhobar$ and $\etabar$, keeping all the
constraints. This has been done in section \ref{sec:newphys}, using the model already considered in \cite{ref:bello}.

\section{The measurement of the parameter $A$.}
\label{sec:A}

The value of the parameter $A$ is obtained from measurements of $\Vcb$ in exclusive and inclusive semileptonic decays of
B hadrons. By definition:
\begin{equation}
\Vcb~=~A~ \lambda^2,~{\rm with}~\lambda~=~\sin \theta_c.
\end{equation}

\subsection{$\Vcb$ measurement using $\Bbar \rightarrow \Dstar \ell^- \nubar$ decays.}

In this channel, $\Vcb$ is obtained by measuring the differential decay rate $\frac{d \Gamma_{D^{\ast}}}{d q^2}$
at maximum value of $q^2$ \cite{ref:neubertvcb1}.
$q^2$ is the mass of the charged lepton-neutrino system. At 
$q^2~=~q^2_{max}$, the $\Dstar$ is produced at rest in the B rest frame and HQET can be invoked
to obtain the value of the corresponding form factor: $F_{D^{\ast}}(w=1)$. The variable $w$ is usually
introduced; it is the product of the 4-velocities of the B and $\Dstar$ mesons:
\begin{equation} 
w~=~v_B \cdot v_{D^\ast}~=~\frac{m_B^2+m_{D^\ast}^2-q^2}{2 m_B m_{D^\ast}},~w=1~{\rm for}~q^2=q^2_{max}.
\label{eq:2.01}
\end{equation}
In terms of $w$, the differential decay rate can be written:
\begin{equation} 
\frac{d BR_{D^{\ast}}}{d q^2}~=~\frac{1}{\tau_{B^0_d}} \frac{G_F^2}{48 \pi^3}m^3_{D^{\ast}}(m_B-m_{D^{\ast}})^2
K(w) \sqrt{w^2-1} F_{D^{\ast}}^2(w) \Vcb^2
\label{eq:2.02}
\end{equation}
in which $K(w)$ is a kinematic factor.

As the decay rate is zero for $w=1$, the $w$ dependence has to be adjusted over the measured range, 
using the previous expression. The form factor is expected to  be $w$ dependent:
\begin{equation} 
F_{D^{*}}(w)~=~F_{D^{*}}(1) [ 1- {\hat{\rho}}^2 (w-1) + {\hat{c}} (w-1)^2 +\emptyset (w-1)^3]
\label{eq:2.03}
\end{equation}

The value of the slope at the origin, ${\hat{\rho}}^2$, is usually also fitted on data and the curvature, ${\hat{c}}$
is neglected or related to  ${\hat{\rho}}^2$, using constraints on the $q^2$ dependence of form factors
which are due to analyticity properties of QCD spectral functions and unitarity \cite{ref:neubertvcb2}:
\begin{equation} 
{\hat{c}}~=~0.66~{\hat{\rho}}^2 -0.11 +\emptyset (\Lambda/m_b)
\label{eq:2.04}
\end{equation} 
At LEP the $q^2$ of the reaction is obtained from the measured characteristics of the final state
in the decay $\Bdb \rightarrow \Dstarp \ell^- \nubar$.
The B meson direction is measured using the position of the primary and of the B decay vertices. The neutrino energy
is obtained by imposing a global energy-momentum conservation on the whole event to define
the energy in each hemisphere and then subtracting the 4-momenta of the measured charged and 
neutral particles. A typical resolution of 3 GeV is achieved on $E_{\nu}$. The $\Dstarp$
has been exclusively reconstructed \cite{ref:alephvcb} \cite{ref:delphivcb} \cite{ref:opalvcb}
or simply the $\pis$ from its cascade decay has been used \cite{ref:delphivcb}. In the first case, resolutions of 0.07 to 0.1
have been obtained on $w$ whereas, in the other, larger samples of events can be analyzed
but with a reduced resolution. In practice, the inclusive analysis does not suffer from larger systematic uncertainties.
Experiments have fitted simultaneously $F_{D^{*}}(1) \Vcb$ and ${\hat{\rho}}^2$ (Table \ref{tab:c12}). 
Results have been combined, including
the CLEOII measurement \cite{ref:cleovcb}, taking into account common systematics among LEP
and CLEO: $BR(\Do \rightarrow \Km \pi^+)$, $\tau_{B^0_d}$, $BR(\Dstarp \rightarrow \Do \pi^+)$ and $\Dstarstar$
rate in semileptonic decays and also additional common systematics between LEP experiments as the branching
fraction of the $\Zz$ to $\bbbar$ pairs and the fraction of $\Bd$ mesons in $b$ jets. Values given in Table \ref{tab:c12}
have been updated using 
common values for all external parameters and include
the recent improvements on the uncertainties on these parameters (see Table \ref{tab:constants}).
At present the dominant systematic
uncertainty ($4 \%$ on $\Vcb$) comes from the error on the $\Dstarstar$ rate in B semileptonic decays.

\begin{table}[htb]
\begin{center}
  \begin{tabular}{|c|c|c|}
    \hline
 Experiment    &  $(F_{D^*}(1) \Vcb)~\times 10^3$   & ${\hat{\rho}}^2$ \\
    \hline
 ALEPH         & $31.7 \pm 1.8 \pm 1.9 $     & $ 0.31 \pm 0.17 \pm 0.08 $   \\
 DELPHI        & $35.7 \pm 2.0 \pm 2.4 $     & $ 0.74 \pm 0.20 \pm 0.17 $   \\
 OPAL          & $32.5 \pm 1.9 \pm 2.2 $     & $ 0.55 \pm 0.24 \pm 0.05 $   \\
    \hline
 LEP average & $33.0 \pm 1.3 \pm 1.8 $     & $   $\\
    \hline
 CLEOII        & $35.1 \pm 1.9 \pm 1.9 $     & $ 0.80 \pm 0.17 \pm 0.17 $   \\
    \hline
 global average& $33.72 \pm 1.25 \pm 1.56 $  & $   $   \\
    \hline
  \end{tabular}
  \caption[]{\it {Measurements of $F_{D^*}(1) \Vcb$ in $\Bbar \rightarrow \Dstar \ell \nubar$.}}
  \label{tab:c12}
\end{center}
\end{table}
\noindent
Table \ref{tab:c12} shows that the individual LEP experiments have the same sensitivity as published results from CLEO.
Using  $F_{D^*}(1)=0.91 \pm 0.06$ \cite{ref:bigivcb}, the following value
is obtained for $\Vcb$:
\begin{equation} 
\Vcb (exclusives)~=~(37.0 \pm 2.2 \pm 2.4)\times 10^{-3}
\label{eq:2.05}
\end{equation}

\begin{table}[htb]
\begin{center}
  \begin{tabular}{|c|c|c|}
    \hline
 Parameter &  Value     &    Reference \\
    \hline
 ${\cal P}(b \rightarrow \Bd)$ & $(39.5 ^{+1.6}_{-2.0})\% $     & \cite{ref:osciw}   \\
 ${\rm BR}(\Do \rightarrow \Km \pi^+)$ & $ (3.83 \pm 0.12)\%$     & \cite{ref:pdg96}     \\
 $\tau_{B^0_d}$ & $ (1.57  \pm 0.04) ps $     & \cite{ref:taugroup}   \\
 ${\rm BR}(\Bd \rightarrow \ell X)$    & $(10.2 \pm 0.5)\%$ & (see below) \\
    \hline
  \end{tabular}
  \caption[]{\it {Values of the parameters used in the present determination of $\Vcb$.
The semileptonic branching fraction for the $\Bd$ meson has been obtained using the inclusive 
semileptonic branching fraction measurement done at the $\Upsilon$(4S) \cite{ref:brcleo} and correcting for the
contribution of charged B mesons by taking into account the difference between $\Bd$
and $\Bm$ lifetimes.}}
  \label{tab:constants}
\end{center}
\end{table} 

\subsection{$\Vcb$ measurement using inclusive semileptonic decays.}

The expression relating the value of $\Vcb$ and the values of the inclusive $b$ lifetime and
semileptonic branching fraction can be found in \cite{ref:bigivcb}:
\begin{equation} 
\Vcb~=~0.0419 \sqrt{\frac{BR(B \rightarrow X_c \ell \nubar)}{0.105}} \sqrt{\frac{1.55}{\tau_B}}
(1 \pm 0.015 \pm 0.010 \pm 0.012)
\label{eq:2.06}
\end{equation} 
The numbers given at the end of this expression reflect different sources of theoretical uncertainties
and have been added linearly in the following. The first uncertainty depends on the evaluation of 
$\alpha_s^{\overline{MS}}$(1 GeV), the second error is related to the uncertainty on the value of $m_b$ 
and the last one reflects uncertainties on the $1/m_Q^3$ and higher power corrections.

\begin{equation} 
\Vcb (inclusives)~=~(41.0 \pm 1.1 \pm 1.5)\times 10^{-3}
\label{eq:2.07}
\end{equation} 
\subsection{Summary on $\Vcb$ measurements.}

The two results obtained for $\Vcb$ have largely uncorrelated experimental and theoretical uncertainties
and are compatible, their average is:
\begin{equation} 
\Vcb ~=~(40.0 \pm 1.6)\times 10^{-3}
\label{eq:2.08}
\end{equation} 
which implies:
\begin{equation} 
A = 0.823 \pm 0.033
\label{eq:aaa}
\end{equation} 
If, instead of adding linearly theoretical errors, their quadratic sum is used, the two 
measurements of $\Vcb$ are marginally compatible within the quoted errors. Applying the
PDG recipe to scale the errors \cite{ref:pdg96}, exactly the same result is obtained: $(40.1 \pm 1.6)\times 10^{-3}$.

This value of $\Vcb$ may appear to be precisely known (4$\%$ relative error) but, experimentally,
only rates which are proportional to $\Vcb^2$ are measured and, for rates, the relative error
becomes 8$\%$. There are thus good prospects to improve further on the precision of $\Vcb$ but this requires also
a good control of theoretical errors for exclusive and inclusive channels.

In the following analysis, the effect of an uncertainty on $\Vcb$ of $\pm 2 \times 10^{-3}$
has been also examined.

\section{Improvements in the determination of the other parameters.}
\label{sec:dms}
\subsection{Present limit on {$\Delta m_s$}.}

A new limit on $\Delta m_s$, $\Delta m_s > 10.2 ps^{-1} ~{\rm at}~95 \% C.L.$,
has been provided by the "B Oscillation Working Group" 
\cite{ref:osciw}. The sensitivity 
of present measurements is at $13.0 ps^{-1}$. 
The definition of the sensitivity and the way the information on $\Delta m_s$ is used in the constraints
have been explained in \cite{ref:bello}.

\subsection{Present value of { $f_B$}.}
\label{sec:fb}
$f_B$ is evaluated from the measurements of $f_{D_{s}}$ and using the extrapolation from the D to the B 
sector as predicted by lattice QCD. More details can be found also in \cite{ref:bello}. 
The value of $f_{D_{s}}$ is deduced 
from the measurements of the branching fractions : $D^+_s \rightarrow \tau^{+} \nu_{\tau}$ and
$D^+_s \rightarrow \mu^{+} \nu_{\mu}$. The different determinations of  $f_{D_{s}}$ \cite{ref:dstaunu} 
result in the following average:

\begin{equation}
f_{D_{s}} = ( 243 \pm 36 )~ MeV
\label{eq:fds}
\end{equation}

In a recent publication, from lattice QCD \cite{ref:milc}, the ratio between the $\Ds$
and the $\Bd$ decay constants has been evaluated:
\begin{equation}
\frac{f_{B_d}}{f_{D_{s}}} = 0.76 \pm 0.07 
\label{eq:fbfds}
\end{equation}

It results :
\begin{equation}
f_{B} = (185 \pm 25(exp.) \pm 17 (theo.)) MeV
\label{eq:fbded}
\end{equation}

\noindent
and using ${B_{B_d}} = 1.36 \pm 0.16$ \cite{ref:bello} it follows :

\begin{equation}
f_{B_d} \sqrt{B_{B_d}}~=~(220 \pm 40)~MeV
\label{eq:fbsqrtb}
\end{equation}

This value is well compatible with the absolute prediction from
lattice QCD ($f_{B_d} \sqrt{B_{B_d}}~=~(200 \pm 50)~MeV$, which was used in our previous analysis ) 
and also with the value favoured by the measurements of 
$ \mid \epsilon_K \mid $, $\Delta m_d$, $\frac{\left | V_{ub} \right |}{\left | V_{cb} \right |}$ and the limit on $\Delta m_s$, 
presented in section \ref{sec:param}.

\subsection {Present value for $\xi$.}  
\label{sec:fbsofbd}
Significant improvements have been achieved in the determination of the $\xi$ parameter and several authors agree 
on a relative precision better than 10$\%$ (see Table \ref{tab:aa}). 

\begin{table}[hbt]
\begin{center}
\begin{tabular}{|c|c|}
\hline
 reference & value \\
\hline
 \cite{ref:milc}    &  $1.10 \pm 0.07 $ \\
 \cite{ref:burino}  &  $1.17 \pm 0.03 $ \\
 \cite{ref:sach}    &  $1.14 \pm 0.08 $ \\ \hline  
\end{tabular}
\caption[]{ \it {Values of the parameter $\xi$ provided by different collaborations
working on lattice QCD. Only recent results have been reported. }}
\label{tab:aa}
\end{center}
\end{table}

The value from reference \cite{ref:milc} has been used in the following.

\subsection{Present value of $B_K$.}
The central value and the uncertainty used for the scaled invariant parameter $B_K$
have not been modified as compared to our previous analysis ($B_K = 0.90 \pm 0.09 $ \cite{ref:bello} ).
These values are in agreement with recent results obtained 
by lattice QCD collaborations recently reported \cite{ref:gupta}:
\begin{equation}
B_K = 0.86 \pm 0.09
\end{equation} 

\section{Results with present measurements. }
\label{sec:mesures}

The region of the $(\rhobar,~\etabar)$ plane selected by the measurements of $\epsilonk$,
$\vubovcb$, $\dmd$ and from the limit on $\dms$ has been obtained assuming Gaussian errors
or flat probability distributions \footnote{The considered flat probability distributions have been taken with the same 
variance as the corresponding Gaussian distributions (their half width is equal
to $\sqrt{3}~ \sigma$).}  for the parameters $B_K$, $f_{B_d} \sqrt{B_{B_d}}$ , $\xi$ and $\vubovcb$.

\subsection{Measured values of $\rhobar$ and $\etabar$.}
\label{secs:rhoetameas}
Central values and uncertainties on $\rhobar$ and $\etabar$ are given in Table \ref{tab:rhoeta}.

\begin{table}[hbt]
\begin{center}
\begin{tabular}{|c|c|c|}
\hline
 fit conditions & $\rhobar$ & $\etabar$ \\
\hline

 Gaussian errors &  $  0.156 \pm 0.090 $  &  $  0.328 \pm 0.054$ \\
 flat errors     &  $ 0.166 ^{+0.109}_{-0.111}  $  &  $ 0.334 \pm {0.057} $ \\
 Gaussian errors, $\sigma(A)=0.04$  &  $ 0.160 \pm 0.094 $  &  $ 0.341 \pm 0.058 $ \\
\hline
\end{tabular}
\caption[]{ \it {Measured values of the parameters $\rhobar$ and $\etabar$. }}
\label{tab:rhoeta}
\end{center}
\end{table}

Contrary to what is usually claimed \cite{ref:concurrence}:
\begin{itemize}
\item the use of flat distributions for the quantites in which systematic errors are dominant,
does not change significantly the results.
\item the allowed region for $\rhobar$ is not symmetric around zero, negative values for $\rhobar$ being clearly
disfavoured:
\begin{equation}
{\cal P}^{Gauss}_{\rho<0}=  6.7 \%,~{\cal P}^{flat}_{\rho<0}=  9.9\%.
\end{equation}
\end{itemize}

\begin{figure}
\begin{center}
{\epsfig{figure=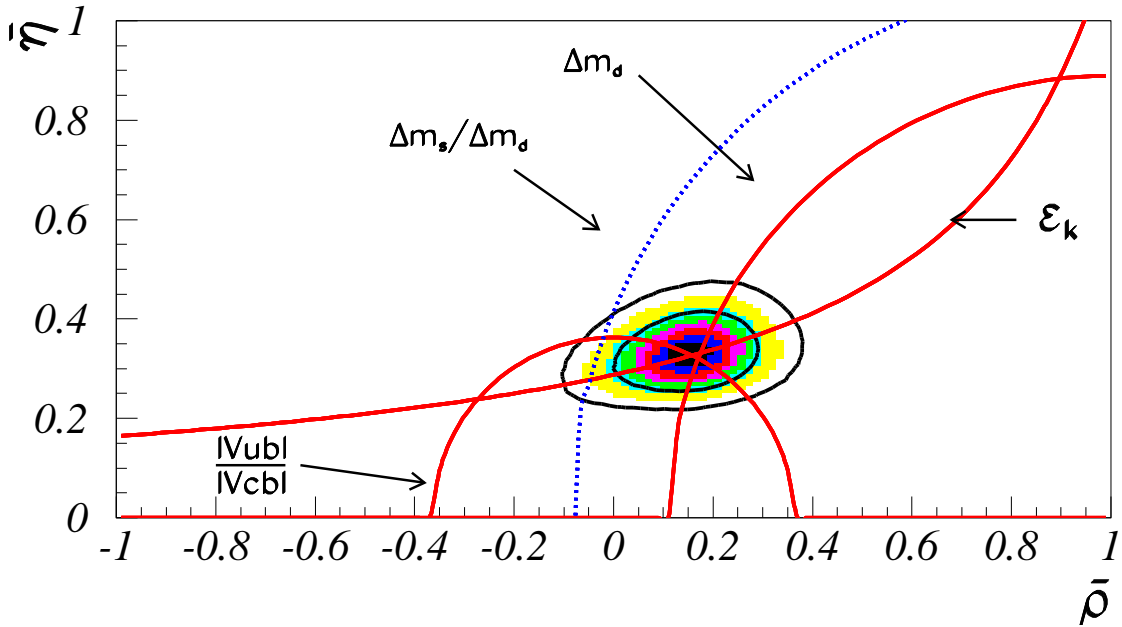,bbllx=30pt,bburx=503pt,bblly=8pt,bbury=249pt,height=8cm}}
\caption{ \it{ The allowed region for $\overline{\rho}$ and $\overline{\eta}$ using the parameters listed in Table \ref{tab:a}.
The contours at 68 $\%$ and 95 $\%$ are shown. The full lines correspond to the central values of the constraints given by
the measurements of  $  \frac{\left | V_{ub} \right |}{\left | V_{cb} \right |} $, $  \mid \epsilon_K  \mid $ and $\Delta m_d$.
The dotted curve corresponds to the 95 $\%$ C.L. upper limit obtained from the experimental limit on
$\Delta m_s$.  }}
\label{fig:haricot2_1}
\end{center}
\end{figure}

The contours corresponding to 68 $\%$ and 95 $\%$ confidence levels are shown in Figure~\ref{fig:haricot2_1}.

\subsection{ Measured values of $\sin 2\alpha$ and $\sin 2\beta$.}

%Present measurements of $\Delta m_d$, $\left | V_{ub} \right |$, $ \mid \epsilon_K \mid $ and the limit on $\Delta m_s$ 
%can be used also to define the allowed region in the plane ($sin2\alpha, sin2\beta$).
%With present measurements the following values have been obtained (Figure \ref{fig:haricot2_seni}):

It is of interest to determine the central values and the uncertainties on the quantities
$\sin 2\alpha$ and $\sin 2\beta$ which can be measured directly at future facilities like 
HERA-B and B factories. Results have been summarized in Table \ref{tab:alphabeta}.

\begin{figure}
\begin{center}
{\epsfig{figure=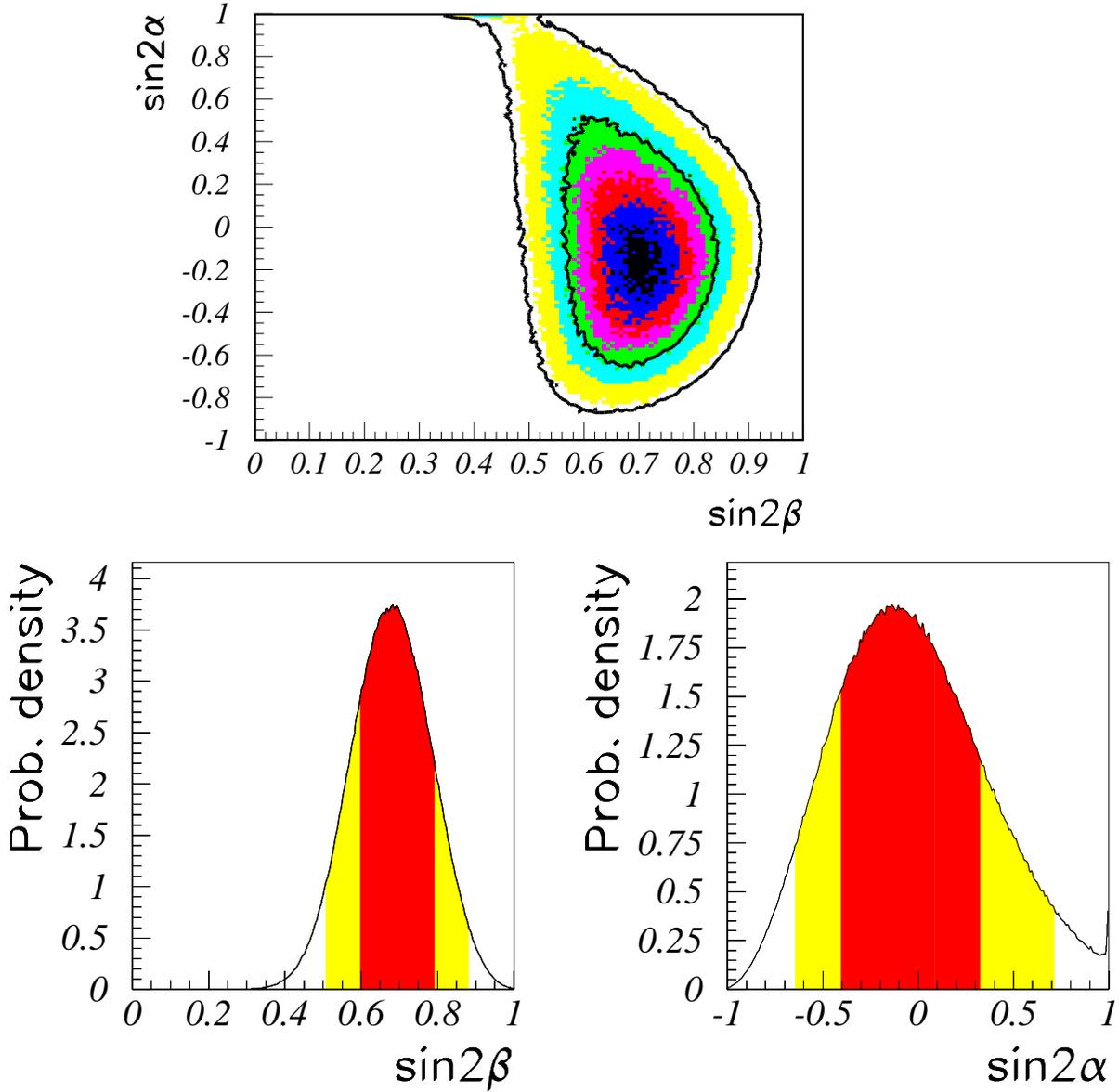,bbllx=0pt,bburx=620pt,bblly=0pt,bbury=521pt,height=16cm}}
\caption{ \it
{ The $sin 2 \alpha$ and $sin 2 \beta$ distributions have been obtained using the contraints corresponding 
to the values of the parameters listed in Table \ref{tab:a}. The contours at 68$\%$ and 95 $\%$ are shown.
The sin2$\alpha$ and sin2$\beta$ distributions are also shown. The dark-shaded and the clear-shaded intervals correspond, 
respectively, to 68$\%$ and 95 $\%$ confidence level regions. }}
\label{fig:haricot2_seni}
\end{center}
\end{figure}

\begin{table}[hbt]
\begin{center}
\begin{tabular}{|c|c|c|}
\hline
 fit conditions & $\sin 2\alpha$ & $\sin 2\beta$ \\
\hline
 Gaussian errors &  $ -0.10 \pm 0.40   $  &  $  0.68 \pm 0.10 $ \\
 flat errors     &  $ -0.10 \pm 0.45  $  & 0.78 $ ^{+0.08}_{-0.16}$ \\
 Gaussian errors, $\sigma(A)=0.04$  &  $ 0.02 \pm 0.43 $  &  $ 0.69 \pm 0.11 $ \\
\hline
\end{tabular}
\caption[]{ \it {Measured values of the parameters $\sin 2\alpha$ and $\sin 2\beta$. }}
\label{tab:alphabeta}
\end{center}
\end{table}
Determinations of these angles, prior to the present analysis can be found in \cite{ref:alphabeta_old}.
Figure \ref{fig:haricot2_seni} gives the correlation between the measurements of these two quantities
and the contours at 68$\%$ and 95 $\%$ C.L..

\subsubsection{Measurement of $sin 2 \beta$.}
The value of $sin 2 \beta$ is rather precisely determined, with an accuracy already at a level expected after the first years of
running at B factories. The situation will improve in the coming years with better
measurements of $\Vcb$, with a possible improvement of the sensitivity of LEP analyses on $\Delta m_s$ and with 
an expected progress from lattice QCD.

\subsubsection{Measurement of $sin 2 \alpha$.}

In our previous analysis \cite{ref:bello}, it was concluded that there was no restriction on the domain
of variation of $sin 2 \alpha$ between -1 and +1. The present study, see Figure \ref{fig:haricot2_seni},
allows to identify a favoured domain for this parameter which is around zero. 
%This result gives support
%to the ansatz of \cite{ref:friche} which predicts that $sin 2 \alpha \simeq 0.$, the CKM unitarity
%triangle being rectangular in $\alpha$.

\subsubsection{Measurement of the angle $\gamma$.}
 It has been proposed in \cite{ref:flman} to restrict the range of variation of the angle $\gamma$
using the measurement of the ratio, $R_1$, of the branching fractions of charged and neutral B mesons
into ${\rm K} \pi$ final states. In the hypothesis that this ratio is below unity, the following
constraint has to be satisfied:
\begin{equation}
\sin ^2 \gamma < R_1
\end{equation}

\begin{figure}
\begin{center}
{\epsfig{figure=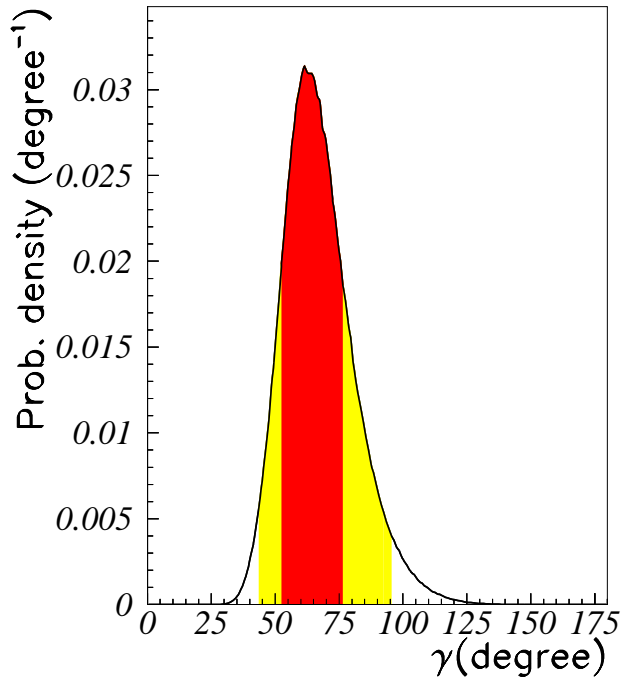,bbllx=0pt,bburx=470pt,bblly=0pt,bbury=515pt,height=9cm}}
\caption{ \it{ The $\gamma$ angle distribution obtained using the same constraints as in Figure \ref{fig:haricot2_1}.
The dark-shaded and the clear-shaded intervals correspond to 68$\%$ and 95 $\%$ confidence level regions respectively.  }}
\label{fig:haricot2_gamma}
\end{center}
\end{figure}

The present result from CLEO \cite{ref:cleokpi}:
\begin{equation}
R_1~=~\fleisher~=0.65 \pm 0.40,
\end{equation}
has a too large uncertainty to be really constraining on $\gamma$. This bound excludes a region
which is symmetric around $\gamma = 90^{\circ}$. 
In fact, as explained already in section \ref{secs:rhoetameas},
negative values of $\rhobar$ are already excluded and the region around $\gamma = 90^{\circ}$ has
a low probability. These restrictions are clearly apparent in Figure \ref{fig:haricot2_gamma} which 
gives the expected density probability distribution for the angle $\gamma$, which is determined to 
be : $\gamma = (64 \pm 12) ^{o}$

\begin{figure}
\begin{center}
{\epsfig{figure=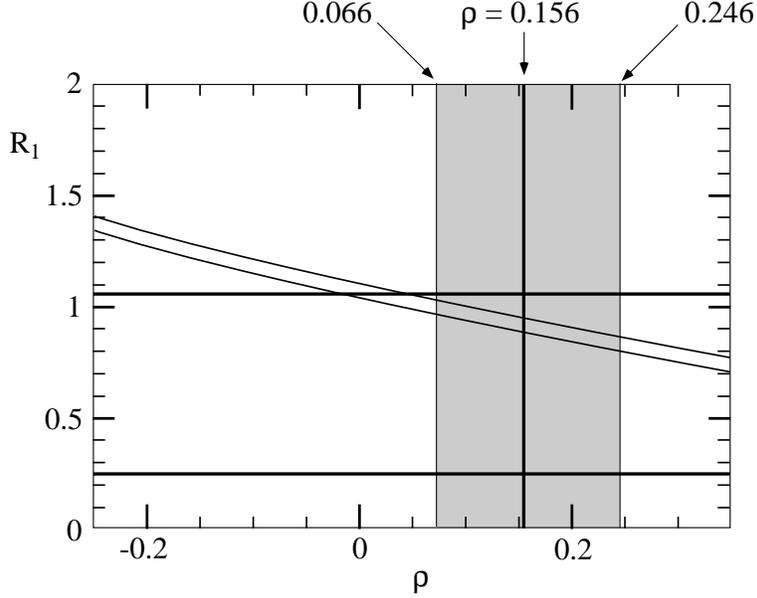,bbllx=76pt,bburx=413pt,bblly=80pt,bbury=355pt,height=8cm}}
\caption{ \it{ The ratio $R_1 = \fleisher$ as a function of the parameter $\rhobar$ taken from \cite{ref:aligreub},
for $\etabar=0.25$ (lower curve) and $\etabar=0.52$ (upper curve). The horizontal
thick lines show the CLEO measurement (with $\pm 1 \sigma$ errors). The shaded vertical
band corresponds to the $\pm 1 \sigma$ interval for $\rhobar$ obtained in the present analysis.}}
\label{fig:alietal}
\end{center}
\end{figure}

At present, theorists do not agree on the effects of hadronic interactions on this analysis
\cite{ref:bkpi}.
But, considering that these effects are under control, the needed
experimental accuracy on $R_1$ has been evaluated such that this measurement provides
an information on $\rhobar$ of similar precision as the one obtained at present.
The model of \cite{ref:aligreub} has been used in which the 
authors have studied the variation of $R_1$
with the $\rhobar$ parameter (Figure \ref{fig:alietal}). The present determination of $\rhobar$ 
corresponds to:

\begin{equation}
R_1~=~\fleisher~=0.89 \pm 0.08~({\rm Gaussian~errors}) .
\end{equation}

\noindent
This result indicates the precision which is needed for a direct measurement of the quantity $R_1$ to reduce the present 
error on the angle $\gamma$ ( and similarly on the parameter $\overline{\rho}$ ).

\subsection{Measurement of the angles $\theta$, $\theta_u$, $\theta_d$ and $\phi$.}
There are nine possibilities to introduce the CP violation phase into the elements of the CKM matrix \cite{ref:friche}.
The authors of \cite{ref:friche} have argued for a parametrization, based on the observed hierarchy in the values 
of quark masses. Several theoretical works \cite{ref:barbieri} show that the observed pattern of fermion masses and mixing 
angles could originate from unified theories with an U(2) flavour symmetry.
The authors of \cite{ref:friche} have introduced four angles which have simple physical interpretations. 
$\theta$ corresponds to the mixing between the families 2 and 3. $\theta_{u(d)}$ is the mixing angle between families
1 and 2, in the up(down) sectors. Finally $\phi$ is responsible for CP
violation and appears only in the elements of the C.K.M. matrix relating the first two families.

This parametrization is given below:

\begin{equation}
\begin{array}{ccc}
V_{CKM} =
&
\left ( \begin{array}{cccc}
s_u s_d c+ c_u c_d e^{-i\phi}~~~~~~~~~~s_uc_dc - c_us_d e^{-i\phi} ~~~     s_us ~~\\
c_us_dc - s_uc_d  e^{-i\phi}  ~~~~~~~~~~    c_uc_dc + s_us_d e^{-i\phi}~~~   c_us ~~\\
-s_d s   ~~~~~~~~~~~~~~~~~~~~~~   -c_ds ~~~~~~~~~~~~~~~~~  c
\end{array} \right )
\end{array}
\end{equation}
where $c_x$ and $s_x$ stand for $cos \theta_x$ and  $sin \theta_x$ respectively.

The four angles are related to the modulus of the following C.K.M. elements:
\begin{equation}
\sin{\theta} = \Vcb \sqrt{1 + \vubovcb^2}
\end{equation}
\begin{equation}
\tan{\theta_u} = \vubovcb
\end{equation}
\begin{equation}
\tan{\theta_d} = \vtdovts
\end{equation}
and
\begin{equation}
\phi= \arccos \left ( \frac {\sin^2{\theta_u} \cos^2{\theta_d} \cos^2{\theta} +
\cos^2{\theta_u} \sin^2{\theta_d} -\left | {\rm V}_{us} \right |^2}
{2 \sin{\theta_u}\cos{\theta_u}\sin{\theta_d}\cos{\theta_d}\cos{\theta} } \right )
\label{eq:phi1}
\end{equation}

The first three equations illustrate the direct relation between the angles $\theta$,
$\theta_u$ and $\theta_d$ and the measurements of B decay and oscillation parameters.

The angle $\phi$ has also a nice interpretation because, in the limit of $\theta=0$ (in practice
$\theta \simeq 2^{\circ}$), the elements $ {\rm V}_{us} $ 
and ${\rm V}_{cd} $ have the same modulus, equal to $\sin{\theta_c}$ 
(${\theta_c}$ is the Cabibbo angle) and can be represented in a complex plane
by the sum of two vectors, of respective lengths $\sin{\theta_u} \cos{\theta_d}$
and  $\sin{\theta_d} \cos{\theta_u}$, making a relative angle $\phi$.
It can be shown that this triangle is congruent to the usual unitarity triangle
\cite{ref:friche} and that $\phi \simeq \alpha$.

\begin{figure}
\begin{center}
{\epsfig{figure=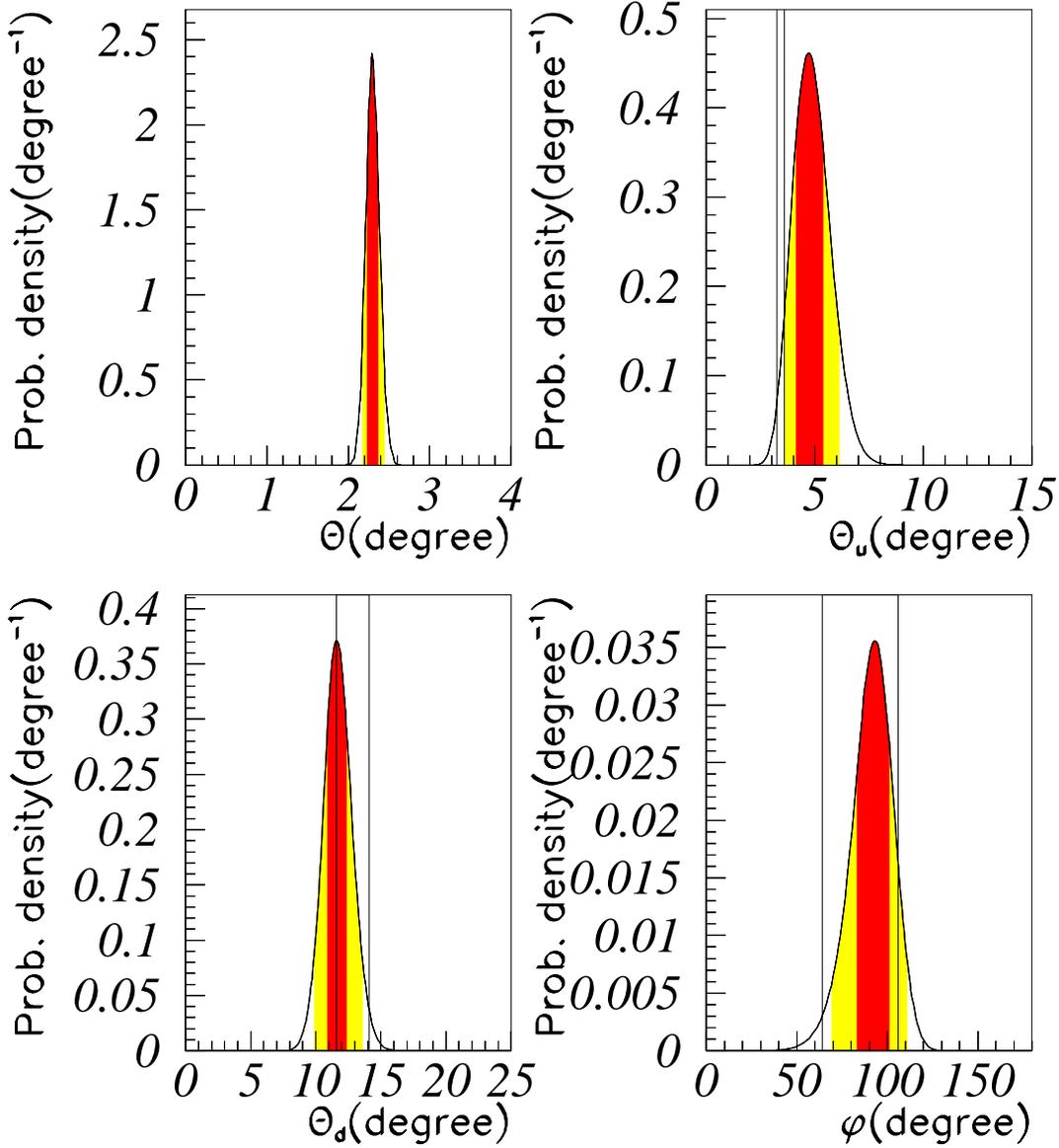,bbllx=0pt,bburx=538pt,bblly=0pt,bbury=519pt,height=16cm}}
\caption{ \it {  The distributions of the angles $\theta$, $\theta_u$, $\theta_d$ and $\phi$ proposed in the 
parametrisation \cite{ref:friche}. The dark-shaded and the clear-shaded intervals correspond to 68$\%$ and 95 $\%$ confidence level
regions respectively. The lines represent the $\pm$1 $\sigma$ region corresponding to the values of the angles obtained using the 
values for the quark masses given in \cite{ref:quark_mass}. }}
\label{fig:friche}
\end{center}
\end{figure}

Using the constraints defined previously, the respective probability distributions
for the four angles have been given in Figure \ref{fig:friche}, and fitted values
are summarized in Table \ref{tab:tabmass}.
\begin{table}[htb]
\begin{center}
\begin{tabular}{|c|c|c|}
\hline
 Angle      &         measured value               &   value expected from quark masses \cite{ref:quark_mass}    \\ \hline
$\theta$    &   (2.30  $\pm$ 0.09)$^{\circ}$       &                                                              \\
$\theta_u$  &   (4.87  $\pm$ 0.86)$^{\circ}$       &     (3.36  $\pm$ 0.35)$^{\circ}$                           \\  
$\theta_d$  &   (11.71 $\pm$ 1.09)$^{\circ}$       &     (12.84 $\pm$ 1.27)$^{\circ}$                          \\  
$\phi$      &    (91.1 $\pm$ 11.8)$^{\circ}$       &     (85    $\pm$ 21)$^{\circ}$                            \\
\hline
\end{tabular}
\caption[]{ \it { Fitted values for the angles of the parametrization \cite{ref:friche}, 
compared with those obtained using the values of the quark masses as given in 
\cite{ref:quark_mass} evaluated at Q$^2$=M$_W^2$ ( the values used for the quark masses are: 
$m_u = 2.35^{+0.42}_{-0.45} MeV,~m_d = 4.73^{+0.61}_{-0.67} MeV,~m_s = 94.2^{+11.9}_{-13.1} MeV~and~m_c = 684^{+56}_{-61} MeV$ ) }}
\label{tab:tabmass}
\end{center}
\end{table}

Another interesting aspect of this parametrization of the C.K.M. matrix is its
possible interpretation in terms of quark masses \cite{ref:friche2}:
\begin{equation}
\tan{\theta_u}=\sqrt{\frac{m_u}{m_c}},~\tan{\theta_d}=\sqrt{\frac{m_d}{m_s}}.
\label{eq:phi2}
\end{equation}

Using the values for the quark masses given in \cite{ref:quark_mass} evaluated at Q$^2$=M$_W^2$,
the values for the angles $\theta_u$ and $\theta_d$ are given 
in Table \ref{tab:tabmass}.
In this interpretation, the angle $\phi$ can be obtained using equations (\ref{eq:phi1}) and (\ref{eq:phi2}).
Present measurements support a value of $\phi$ close to $90^{\circ}$
which corresponds to the maximal CP violation scenario of \cite{ref:friche}.\\
The present analysis indicates that a lower value for $m_u$ is favoured or that the expression relating $\theta_u$with the $u$ and
the $c$ quark masses has to be corrected.

\section{Tests of the internal consistency of the Standard Model for CP violation.}
\label{sec:param}

Four constraints, three measurements and one limit, have been used until now 
to measure the values of the two parameters $\rhobar$ and $\etabar$. It is also possible
to remove, from the fit, the external information on the value of one of the constraints or of another
parameter entering into the Standard Model expressions for the constraints. Each of these 
quantities will be considered, in turn, and fitted in conjunction with $\rhobar$ and $\etabar$.
The results will have some dependence in the central values taken for all the other parameters
but, the main point in this study, is to compare the uncertainty on a given quantity determined in this way
to its present experimental or theoretical error. This comparison allows to quantify the importance
of present measurements of the different quantities in the definition of the allowed region
in the $(\rhobar,~\etabar$) plane.
Results have been summarized in Table \ref{tab:b}.

\begin{table}[htb]
\begin{center}
\begin{tabular}{|c|c|c|c|}
\hline
 parameter              & Fitted value                & Fitted value                   & Present value \\
                        & (Gaussian errors)           & (flat error dist.)             &               \\
\hline 
  $\Delta m_s$          &  $(12 \pm 4) ~ps^{-1}$        &  $(9.5^{+5.3}_{-3.0}) ~ps^{-1}$  & $>$ 10.2 ps$^{-1} ~{\rm at}~95 \% ~C.L.$ \\

  $\vubovcb$            & $ 0.085^{+0.037}_{-0.023}$  &  $0.090^{+0.045}_{-0.026}$     & $0.08 \pm 0.02 $   \\

    $B_K$               &  $ 0.82^{+0.45}_{-0.24}$   &  $ 0.83^{+0.47}_{-0.26}$      & $ 0.90 \pm 0.09$    \\

$f_{B_d}\sqrt{B_{B_d}}$ &   ($213^{+21}_{-20}$) MeV   & ($214^{+23}_{-22}$) MeV        & ($220 \pm 40 $) MeV \\

 $\overline{m_t}(m_t)$  &  $(165^{+52}_{-40})$ GeV    & $(166^{+61}_{-42})$ GeV        & $(168 \pm 6)$ GeV  \\

  A                     & $ 0.82^{+0.15}_{-0.08}$     &  $0.81^{+0.17}_{-0.10}$        & $0.823 \pm 0.033$   \\ 
\hline
\end{tabular}
\caption[]{ \it {Fitted values of the different parameters obtained simultaneously with
$\rhobar$ and $\etabar$ after having removed, in turn, their contribution in the different
constraints.}}
\label{tab:b}
\end{center}
\end{table}
 
\subsection{Expected value for the $\Bs \Bsbar$ oscillation parameter, $\dms$.}

\begin{figure}
\begin{center}
{\epsfig{figure=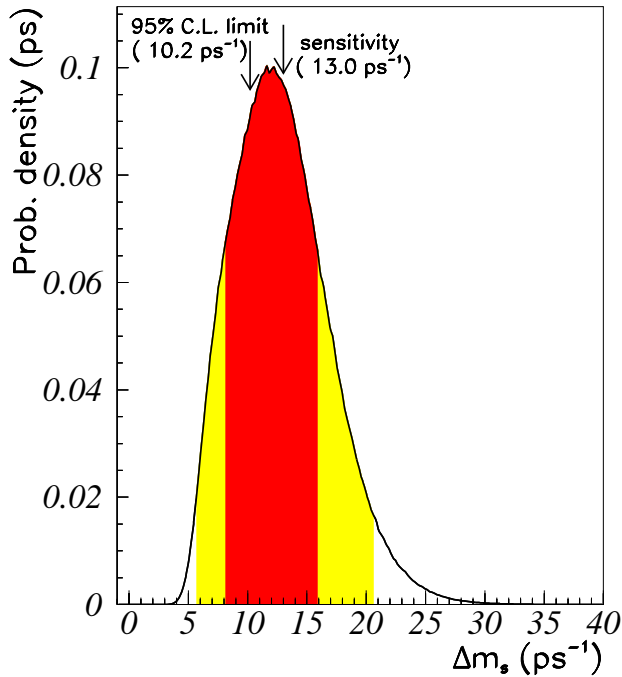,bbllx=0pt,bburx=470pt,bblly=0pt,bbury=515pt,height=9cm}}
\caption{ \it{ The $\Delta m_s$ probability distribution obtained with the same constraints as in Figure \ref{fig:haricot2_1}.
The dark-shaded and the clear-shaded intervals correspond to 68$\%$ and 95 $\%$ confidence level regions respectively. }}
\label{fig:haricot2_dms}
\end{center}
\end{figure}

Removing the constraint from the measured limit on the mass difference between the strange
B meson mass eigenstates, $\dms$, the density probability distribution for $\dms$
is given in Figure \ref{fig:haricot2_dms}.
The present limit excludes already a large fraction of this distribution. Present analyses at LEP
are situated in a high probability region for a positive signal and this is still a challenge
for LEP collaborations.

\subsection{Top mass measurement.}

If the information on the top mass measurement by CDF and D0 collaborations is removed,
the fitted value for $\overline{m_t}(m_t)$ is :  $\overline{m_t}(m_t)$ = $(165^{+52}_{-40})$ GeV.
The present determination of $m_t$ with a $\pm$6 GeV error has thus a large impact on the present analysis.

\subsection{Measurement of $A$.} 

The central value determined for $A$ is close to the direct measurement: A  = $ 0.82^{+0.15}_{-0.08}$ 
Direct measurements of $\Vcb$ are thus important to constrain the allowed region in the $(\rhobar,~\etabar$)
plane if their relative error is below 10$\%$. The density probability distribution for the parameter $A$ is given 
in Figure \ref{fig:haricot2_bk}.

\subsection{Measurement of $\vubovcb$.} 
The central value determined for  $\vubovcb$ is close to the direct measurement. This indirect measurement shows the importance 
of having a precision on $\vubovcb$ better than 30$\%$. The density probability distribution for the parameter $\vubovcb$ 
is given in Figure \ref{fig:haricot2_bk}

\subsection{Measurement of $B_K$.}

The density distribution for the parameter $B_K$ is given in Figure \ref{fig:haricot2_bk}.
It indicates that:
\begin{itemize}
\item values of $B_K$ smaller than 0.6 are excluded at 93$\%$ C.L. (90$\%$ C.L. if flat error distributions are considered),
 
\item large values of $B_K$ are compatible with the other constraints
over a large domain.

\end{itemize}

The present estimate of $B_K$, from lattice QCD, with a 10$\%$ relative error has thus a large
impact for the present analysis.

\subsection {Measurement of $f_B \sqrt{B_B} $. }
\label{sec:523}

\begin{figure}
\begin{center}
{\epsfig{figure=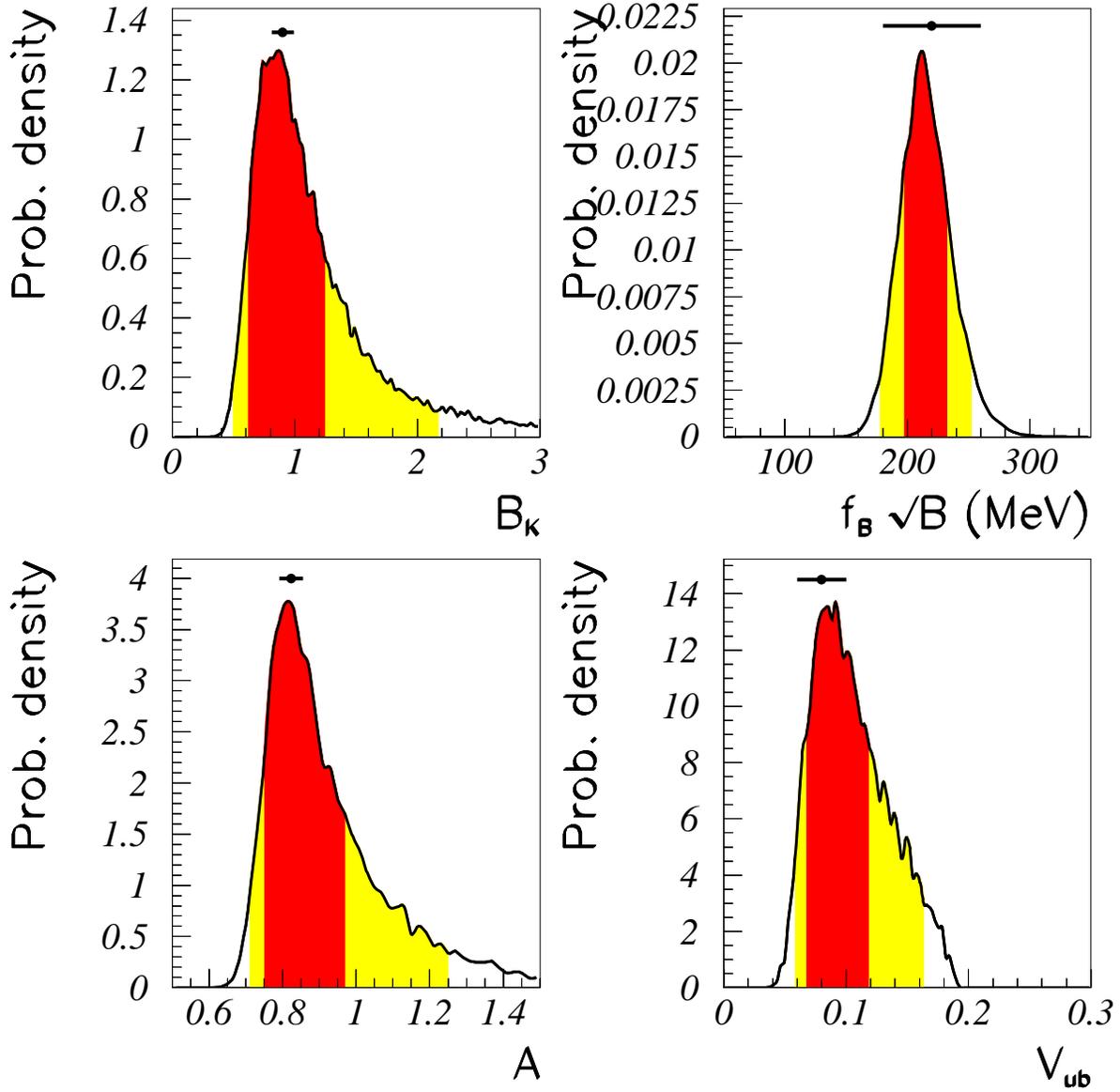,bbllx=0pt,bburx=525pt,bblly=0pt,bbury=525pt,height=16cm}}
\caption{ \it{ The $B_K$, $f_{B_d}\sqrt{B_{B_d}}$, A and $\vubovcb$ probability distributions obtained with the same 
constraints as in Figure \ref{fig:haricot2_1}. The dark-shaded and the clear-shaded intervals correspond to 68$\%$ and 
95 $\%$ confidence level regions respectively. The points and the errors bars show the central values and the uncertainties 
for these parameters used in the present analysis. }}
\label{fig:haricot2_bk}
\end{center}
\end{figure}

A very accurate value is obtained:
\begin{equation}
f_{B_d}\sqrt{B_{B_d}} = \left ( 213 ^{+21}_{-20}\right ) MeV
\label{eq:fb_mes}
\end{equation}
This result is, in practice, in agreement and somewhat more precise than the present evaluation 
of this parameter ( eq. \ref{eq:fbsqrtb} )
The only possibility to obtain another determination of this quantity with a similar or better accuracy,
is to measure $f_{D_s}$ and $f_{D^+}$ at a $\tau/$Charm factory and to use results from lattice QCD
on $\frac{f_{B_d}}{f_{D_s}}$ to deduce the value of $f_{B_d}$ \cite{ref:bello}.
The density probability distribution for the parameter $f_{B_d}\sqrt{B_{B_d}}$ is given in Figure \ref{fig:haricot2_bk}

\section{Limits for a given New Physics scenario.}
\label{sec:newphys}

It has been shown in \cite{ref:bello} that there can be additional contributions to 
$\Delta m_d$ and $ \mid \epsilon_K \mid $ expected to come from the presence of new physics beyond the Standard Model.
The effect of new physics can be parametrized by introducing an extra parameter $\Delta$. 
This analysis is for the moment quite simple since only one parameter is introduced both in the expressions of $\Delta m_d$ 
and of $ \mid \epsilon_K \mid $.
The present data can be then fitted using the new expressions for $\Delta m_d$ and $ \mid \epsilon_K \mid$,
fitting $\Delta$ together with the parameters $\overline{\rho}$ and $\overline{\eta}$, the result is :
$$
\overline{\rho} = 0.147 ^{+0.09}_{-0.12} ~~ ; ~~ \overline{\eta} = 0.328 ^{+0.071}_{-0.070}    \\
$$
\begin{equation}
\Delta = 2.36^{+1.05}_{-0.76}
\label{eq:fitsusy}
\end{equation}
It has to be reminded that the Standard Model predicts: $\Delta=2.55 \pm 0.15 $ ( in the definition
of $\Delta$ the top contribution is included and the error takes into account the uncertainty on the 
top mass). As in \cite{ref:bello} implications of the achieved precision on $\Delta$, can be evaluated 
in a particular  framework of the MSSM extension of the Standard Model in which the stop-right ($\tilde {t}$) and
higgsinos are light, assuming that $tan ^2 \beta$ is lower than $m_t/m_b$ and that the stop-left and the gauginos are heavy.
For this study the contribution from the charged Higgs sector has been neglected and the stop-right and the charged 
higgsino are supposed to have the same mass (generically indicated as $m_{SUSY}$ in the following).
The theoretical framework is discussed in \cite{ref:susy}, results are summarized in Table \ref{tab:susy}.
\begin{table}
\begin{center}
\begin{tabular}{|c|c|c|c|}
\hline
  $\Delta$               & $m_{SUSY}$ , tan $\beta$ = 1 &  tan $\beta$=1.5 &  tan $\beta$ = 5  \\ \hline
 $2.36^{+1.05}_{-0.76}$  &         $>135$               &        $>100$    &   $>80$           \\ \hline
\end{tabular}
\caption[]{ \it {Results on the $\Delta$ parameter. 95 $\%$ C.L limits on $m_{SUSY}$ are given in $GeV/c^{2}$ unit.}}
\label{tab:susy}
\end{center}
\end{table}

Interesting limits on $m_{SUSY}$ can be put for low values of tan $\beta$.

\section{Conclusions.}
The $\overline{\rho}$ and $\overline{\eta}$ parameters have been determined using the constraints from the measurements of
$  \frac{\left | V_{ub} \right |}{\left | V_{cb} \right |} $, $  \mid \epsilon_K  \mid $, $\Delta m_d$ and from the limit on
$\Delta m_s$:
$$
\rhobar=0.156 \pm 0.090  ,  ~\etabar=0.328 \pm 0.054
$$
Contrary to similar studies in this field, which claim a rather symmetric interval of variation for $\rhobar$, 
around zero \cite{ref:concurrence}, the negative $\rhobar$ region is excluded at about 93$\%$ C.L..

Present measurements assume the unitarity of the CKM matrix. In this framework the value of $sin 2\beta$ and $sin 2\alpha$ 
can be also deduced. They are :
$$
\sin 2 \alpha =  -0.10 \pm 0.40    ,~\sin 2 \beta =  0.68 \pm 0.10  
$$
The value of $ sin 2\beta$ has an accuracy similar to the one expected after the first years 
of running at B factories. 
The region centered on zero is favoured for $sin 2\alpha$ in accordance
with the anzats of \cite{ref:friche}.

Values of the angle $\gamma$ larger or equal to $90^{\circ}$ are disfavoured ($\gamma = 64 \pm 12 ^{o}$)
and present determinations of $\rhobar$ and $\etabar$ allow to predict, in a precise way, the value for the
ratio (using the model of \cite{ref:aligreub}):
$$
R_1 = \fleisher = 0.89 \pm 0.08
$$

The other parametrization of the CKM matrix, proposed in \cite{ref:friche} has been studied. 
The four corresponding parameters, which are angles, have been determined:
$$
\theta= (2.30 \pm 0.09 )^{\circ},~\theta_u= (4.87 \pm 0.86)^{\circ}
$$
$$
\theta_d= (11.71 \pm 1.09)^{\circ},~\phi= (91.1 \pm 11.8)^{\circ}
$$
Present values of $\phi$ agree with the maximal CP violation scenario proposed in \cite{ref:friche}.

The internal consistency of the Standard Model expectation for CP violation, expressed by a single
phase parameter in the C.K.M. matrix, has been verified by removing, in turn, the different
constraints imposed by the external parameters. No anomaly has been noticed with respect
to the central values used in the present analysis, in agreement with the small value of the $\chi^2$
of the fit of $\rhobar$ and $\etabar$ alone. This study has mainly quantified the needed accuracy
on the different determinations of these parameters so that they bring useful constraints in the
determination of $\rhobar$ and $\etabar$. In this respect, present uncertainties on $m_t$, $\Vcb$ and
$B_K$ have important contributions.

Low values of $B_K$, below 0.6, are not compatible with the present analysis at 90$\%$ C.L..

$\Delta m_s$ is expected to lie within 1$\sigma$ between 8 and 16 $ps^{-1}$. A measurement of this parameter seems to be possible
at LEP.

More accurate measurements, still expected at CLEO and LEP, and more precise evaluations of non perturbative QCD parameters from
lattice QCD, will improve these results in the coming years, before the start up of B factories.
A Tau/Charm factory providing accurate values for $f_{D^+}$ and $f_{D_s}$ is expected to have important contributions in this
analysis.

\section{Acknowledgments}
We would like to thank C.W. Bernard and R. Gupta for all the useful interactions on the subjects related to non 
perturbative QCD parameters. A special thanks to F. Richard and D. Treille for constant and warm support in this work.

\end{document}